\documentclass[preprint]{aastex}

\begin{document}

\title{The Physical Nature and Orbital Behavior of the Eclipsing System DK Cygni}
\author{Jae Woo Lee$^{1,2}$, Jae-Hyuck Youn$^{1}$, Jang-Ho Park$^{1,3}$, and Marek Wolf$^{4}$}
\affil{$^1$Korea Astronomy and Space Science Institute, Daejeon 305-348, Korea}
\email{jwlee@kasi.re.kr, jhyoon$@$kasi.re.kr, pooh107162@kasi.re.kr}
\affil{$^2$Astronomy and Space Science Major, Korea University of Science and Technology, Daejeon 305-350, Korea}
\affil{$^3$Department of Astronomy and Space Science, Chungbuk National University, Cheongju 361-763, Korea}
\affil{$^4$Astronomical Institute, Faculty of Mathematics and Physics, Charles University in Prague, 180 00 Praha 8, V Hole\v sovi\v ck\'ach 2, Czech Republic}
\email{wolf@cesnet.cz}

\begin{abstract}
New CCD photometry is presented for the hot overcontact binary DK Cyg, together with reasonable explanations for the light and 
period variations. Historical light and velocity curves from 1962 to 2012 were simultaneously analyzed with the Wilson-Devinney
(W-D) synthesis code. The brightness disturbances were satisfactorily modeled by applying a magnetic cool spot on the primary star. 
Based on 261 times of minimum light including 116 new timings and spanning more than 87 yrs, a period study reveals that 
the orbital period has varied due to a periodic oscillation superposed on an upward parabola. The period and semi-amplitude of 
the modulation are about 78.1 yrs and 0.0037 d, respectively. This detail is interpreted as a light-travel-time effect due to 
a circumbinary companion with a minimum mass of $M_3$=0.065 $M_\odot$, within the theoretical limit of $\sim$0.07 M$_\odot$ 
for a brown dwarf star. The observed period increase at a fractional rate of $+$2.74 $\times $10$^{-10}$ is in excellent 
agreement with that calculated from our W-D synthesis. Mass transfer from the secondary to the primary component is mainly 
responsible for the secular period change. We examined the evolutionary status of the DK Cyg system from the absolute dimensions. 
\end{abstract}

{}

\section{INTRODUCTION}

W UMa-type binaries are interesting systems in which both components are in contact with each other and share a common envelope.
They are classified into two subclasses, A and W, defined observationally by Binnendijk (1970). The A-subtype systems are 
those that show primary minima due to eclipses of their larger and more massive components, while the reverse is true for 
W-subtype systems. Also, the A's are statistically hotter and more massive than the W's and have evolved beyond the zero-age 
main sequence (ZAMS), in some cases almost to the terminal-age main sequence (TAMS). The W's are close to the ZAMS. And the A's
typically have much more extreme mass ratios. The overcontact binaries are thought to have evolved from detached binaries via 
angular momentum loss through magnetic braking caused by stellar winds and ultimately to coalesce into single stars
(Bradstreet \& Guinan 1994). In this scenario, the existence of the third components may have played an important role in 
the formation of the initial tidal-locked detached progenitors through energy and angular momentum exchanges 
(Eggleton \& Kisseleva-Eggleton 2006; Fabrycky \& Tremaine 2007). The statistical study by Pribulla \& Rucinski (2006) indicates 
that most W UMa binaries have companions. This suggests that the circumbinary objects are necessary for the formation and 
evolution of short-period binaries.

DK Cyg ($\rm BD+33^{o} 4304$, HIP 106574, TYC 2712-250-1; $V\rm_T$=$+$10.61, $(B-V)\rm_T$=$+$0.45; A8V) was discovered to be 
a W UMa-type variable from photographic observations by Guthnick \& Prager (1927). Prior to 2000, photoelectric light curves  
were made by Binnendijk (1964), Paparo et al. (1985), and Awadalla (1994), wherein the observations of the second paper are 
not currently available. Double-lined radial-velocity (RV) curves were obtained by Rucinski \& Lu (1999). They determined 
the velocity semi-amplitudes of the primary and secondary components to be $K_1$ = 87.89 km s$^{-1}$ and 
$K_2$ = 270.46 km s$^{-1}$, respectively, and classified this system as an A-subtype overcontact binary with a spectral type 
of A8 V. Baran et al. (2004) computed the binary parameters from their photoelectric observations, by fixing the mass ratio of 
$q$=$M_2$/$M_1$=$K_1$/$K_2$=0.306 corrected for proximity effects and considering both a cool spot on the primary star and 
a third light. The results indicate that DK Cyg is an overcontact binary with an orbital inclination of $i$=82$^\circ$.5, 
a temperature difference of $\Delta T$=800 K between the components, a fill-out factor of $f$ = 30 \%, and a third light of 
$l_3$=2$\sim$7 \%. They suggest that the third light source may be a star bound to the eclipsing system or faint stars present 
in the 30$\arcsec$ aperture used in their observations. 

Most recently, Elkhateeb \& Nouh (2014) separately re-analyzed the previously published $V$ light curves, except for that 
of Baran et al. (2004). For the light-curve modeling, the mass ratio of $q$=0.32 by Rucinski \& Lu (1999) was fixed 
throughout the analyses and at least a cool spot to each component was applied. Absolute dimensions were also obtained 
from their photometric elements and from the spectroscopic results of Baran et al. (2004), and they concluded that 
the primary component is located nearly on the ZAMS in both the mass-luminosity (M-L) and mass-radius (M-R) diagrams and
that the secondary is above the TAMS tracks in these diagrams. 

After the orbital period change was first examined by Paparo et al. (1985), the period has been studied by 
Awadalla (1994), Wolf et al. (2000), Borkovits et al. (2005), and Elkhateeb \& Nouh (2014). From a quadratic least-squares 
fit, they all reported that the orbital period is increasing and that its main cause is explained by mass transfer from 
the secondary to the primary component. Nonetheless, the period variation still has not been studied as thoroughly as 
can be desired. Eclipse timings are now long enough to study long-term orbital behavior. In this paper, we present 
improved descriptions of the physical properties of DK Cyg from detailed analyses of the RV and light curves and 
eclipse timings, based on all historical data as well as our new CCD observations.

\section{CCD PHOTOMETRIC OBSERVATIONS}

We carried out CCD photometry of DK Cyg on 9 nights from 2012 September 20 through October 21 in order to obtain new multiband 
light curves. The observations were taken with a PIXIS: 2048B CCD camera and a $BVR$ filter set attached to the 61-cm reflector 
at Sobaeksan Optical Astronomy Observatory (SOAO) in Korea. The instrument and reduction method are the same as those described 
by Lee et al. (2013). TYC 2712-1372-1 (2MASS J21351474+3430533; C) was chosen as a comparison star and no peculiar light variations 
were detected against measurements of two check stars, TYC 2712-1841-1 (2MASS J21350318+3434120; K$_1$) and TYC 2712-1886-1 (K$_2$). 
The reference stars were imaged on the chip at the same time as the program target.

A total of 4,411 individual observations was obtained in the three bandpasses (1,478 in $B$, 1,475 in $V$, and 1,458 in $R$) and 
a sample of them is listed in Table 1. The natural-system light curves are shown in Figure 1 as differential magnitudes 
{\it versus} orbital phases, which were computed according to the ephemeris for our cool-spot model on the primary star described 
in the following section. The (K$_1-$C) magnitude differences in the $B$ band are plotted in the uppermost part of the figure. 
As shown in the figure, the SOAO observations are typical of W UMa type and display light changes at a primary eclipse.
Specifically, those data taken on 2012 September 23 are very different from the other data, and they were excluded from 
our light-curve analysis. The secondary minimum seems to indicate a total eclipse but is distorted and inclined.

In addition to these complete light curves, 13 eclipse timings were observed in both 2013 and 2014 using an ARC 4K CCD camera 
and a $V$ band attached to the 1.0-m reflector at the Mt. Lemmon Optical Astronomy Observatory (LOAO) in Arizona, USA. 
TYC 2712-1372-1 and TYC 2712-1841-1 also served as the comparison and check stars, respectively, for these data collections. 
Details of the LOAO observations have been given previously by Lee et al. (2012).

\section{LIGHT-CURVE SYNTHESIS AND ABSOLUTE DIMENSIONS}

Figure 2 assembles the $BV$ observations obtained from 1962 to 2012 by requiring the maximum lights at the first quadrature 
to be identical. Although historical light curves have not appreciably displayed year-to-year light variability, 
the light maxima (Max I and Max II) are displaced to around phases 0.24 and 0.76, respectively. Such changes may be caused by 
local photospheric inhomogeneities and can be explained by spot activity on the components. In order to obtain a unique solution 
for DK Cyg, three sets of light curves (Binnendijk 1964, Baran et al. 2004, SOAO), after normalization to unit light at 
phase 0.25, were simultaneously modeled with the RV curves of Rucinski \& Lu (1999). The data of Awadalla (1994) were not included 
in our analysis because they diverge from all the others and display very peculiar light curves in the second quadrature. 

For the light-curve synthesis, we used the contact mode 3 of the 2003 version of the Wilson-Devinney binary code 
(Wilson \& Devinney 1971; Wilson 1979, 1990, 1993; Van Hamme \& Wilson 2003; hereafter W-D) and a weighting scheme similar 
to that for the eclipsing systems RU UMi (Lee et al. 2008) and V407 Peg (Lee et al. 2014b). Table 2 lists the RV and 
light-curve sets analyzed in this paper and their standard deviations ($\sigma$). The surface temperature of the hotter and 
more massive primary star was assumed to be $T_{1}$=7,500 K, appropriate for its spectral type A8V given by Rucinski \& Lu (1999).
The bolometric albedos and the gravity-darkening exponents were fixed at standard values of $A$=0.5 and $g$=0.32 for stars 
with common convective envelopes. The logarithmic bolometric ($X$, $Y$) and monochromatic ($x$, $y$) limb-darkening coefficients 
were initialized from the values of van Hamme (1993) in concert with the model atmosphere option. Before the historical curves
of DK Cyg were analyzed, the light-travel time (LTT) effects proposed in the following section were applied to the observed times
of all individual points (Lee et al. 2013): HJD$_{\rm new}$=HJD$_{\rm obs}$--$\tau_{3}$. In this paper, the subscripts 1 and 2 
refer to the primary and secondary stars being eclipsed at Min I (at phase 0.0) and Min II, respectively.

Light variations of close binaries may be due to large cool starspots, to hot regions such as faculae, or to gas streams 
and their impact on a companion star. Because DK Cyg should have a common convective envelope and both components are 
fast-rotating stars, we can apply magnetic cool spots on the component stars. There is, at present, no way to know which 
spot model is more efficient in creating light changes. Thus, a cool spot on either of the components is considered to model 
the light curves. Although it is difficult to distinguish between the two spot models from only the light-curve analysis, 
the cool spot on the primary gives a better fit than that on the secondary component. Final results are given in Table 3 
together with the spot parameters. The synthetic $V$ light curves are plotted as the solid curves in Figure 3, while 
the synthetic RV curves are plotted in Figure 4. As shown in the figures, our spot model describes the historical light curves 
quite well. Finally, to study the spot and luminosity behavior of DK Cyg, we re-analyzed three datasets separately by adjusting 
the orbital ephemeris ($T_0$ and $P$), spot, and luminosity among the light-curve parameters. The results are given in Table 4,
which reveal that the light ratios and most spot parameters have been almost constant with time. In all the procedures that 
have been described, we included as a free parameter a third light but found that the parameter remained zero within its margin 
of error.

From the light and RV parameters, we obtained the absolute dimensions listed in Table 5. The luminosity ($L$) and 
bolometric magnitudes ($M_{\rm bol}$) were obtained by adopting $T_{\rm eff}$$_\odot$=5,780 K and $M_{\rm bol}$$_\odot$=+4.73 
for solar values. The temperature of each component has an error of 200 K in accordance with the unreliability in the spectral 
classification. For the absolute visual magnitudes ($M_{\rm V}$), we used the bolometric corrections (BCs) from the relation 
between $\log T_{\rm eff}$ and BC given by Torres (2010). With an apparent visual magnitude of $V$=+10.57 (H\o g et al. 2000) 
and the interstellar absorption of $A_{\rm V}$=0.65 (Schlegel et al. 1998), we have calculated the distance to the system 
to be 366$\pm$21 pc. This is too large compared with the value 226$\pm$91 pc taken by trigonometric parallax (4.42$\pm$1.78) 
from the Hipparcos and Tycho Catalogues (ESA 1997). The difference may partly result from the large uncertainty of 
the Hipparcos measurements for the DK Cyg system.

\section{ORBITAL PERIOD STUDY}

From our observations, 19 new times of minimum light and their errors were determined with the weighted means for the timings 
in each bandpass by using the method of Kwee \& van Woerden (1956). In addition, 93 eclipses were newly derived by us from 
the WASP (Wide Angle Search for Planets) public archive (Butters et al. 2010) and four timings from the data of 
Baran et al. (2004). For a period study of DK Cyg, 145 eclipse timings (26 visual, 1 photographic, 33 photoelectric and 85 CCD) 
were collected from the data base of Kreiner et al. (2001) and from more recent literature. All photoelectric and CCD timings 
are listed in Table 6, wherein the second column gives the HJED (Heliocentric Julian Ephemeris Date) timings transformed to 
the terrestrial time scale (Bastian 2000). Because many timings of the system have been published without error information, 
the following standard deviations were assigned to timing residuals based on each observational method: $\pm$0.0063 d for visual 
and photographic, and $\pm$0.0013 d for photoelectric and CCD minima. Relative weights were then scaled from the inverse squares 
of these values.

The observed ($O$) $-$ calculated ($C$) residuals from the quadratic ephemeris seem to indicate the existence of 
an additional oscillation producing a small scattering of about $\pm$0.004 d. The periodic variation could be identified as 
an LTT effect caused by the presence of a third body orbiting around the eclipsing pair. Thus, the eclipse timings were fitted 
to a quadratic {\it plus} LTT ephemeris:
\begin{eqnarray}
C = T_0 + PE + AE^2 + \tau_{3},
\end{eqnarray}
where $\tau_{3}$ is the LTT due to a circumbinary companion (Irwin 1952) and includes five parameters ($a_{12}\sin i_3$, $e$, 
$\omega$, $n$ and $T$). Here, $a_{12}\sin i_3$, $e$, and $\omega$ are the orbital parameters of the eclipsing pair around 
the mass center of the triple system. The parameters $n$ and $T$ denote the Keplerian mean motion of the mass center of 
the eclipsing pair and the epoch of its periastron passage, respectively. The Levenberg-Marquardt algorithm (Press et al. 1992) 
was applied to solve for the eight parameters of the ephemeris (Irwin 1959), the results of which are summarized in Table 7, 
together with related quantities. The parameter errors are calculated from the 10,000 Monte Carlo bootstrap-resampling experiments 
following the procedure described by Lee et al. (2014a). The quadratic {\it plus} LTT ephemeris resulted in a smaller 
$\chi^2_{\rm red}$=1.05 than the quadratic ephemeris ($\chi^2_{\rm red}$=1.66). Our absolute dimensions in Table 5 have been 
used for these and subsequent calculations.

The $O$--$C$ diagram constructed with the linear terms of the quadratic {\it plus} LTT ephemeris is plotted in Figure 5. 
The photoelectric and CCD residuals from the complete ephemeris appear as $O$--$C_{\rm full}$ in the fifth column of Table 6. 
As displayed in the figure, the quadratic {\it plus} LTT ephemeris gives a satisfactory fit to the mean trend of the residuals. 
If the third object is on the main sequence and its orbit is coplanar with the eclipsing binary ($i_3 \simeq$ 83$^\circ$), 
the mass of the object is computed to be $M_3$ = 0.065 M$_\odot$ and its radius and temperature are calculated to be 
$R_3$=0.073 R$_\odot$ and $T_3$=3090 K, respectively, using the empirical relations from well-studied eclipsing binaries 
(Southworth 2009). The circumbinary object has a mass within the hydrogen-burning limit of $\sim$0.07 M$_\odot$, making it 
difficult to detect such a companion from the light-curve analysis and spectroscopic observations. 

The quadratic term (A) in Equation (1) indicates a continuous period increase with a rate of d$P$/d$t$ = 
$+$9.99 $\times 10^{-8}$ d yr$^{-1}$, corresponding to a fractional period change of $+$2.74 $\times $10$^{-10}$. 
This value is in excellent agreement with $+$2.69$\times$10$^{-10}$ derived from our W-D synthesis, independently of 
the eclipse timings. The most common explanation of the secular period increase in overcontact systems is a mass transfer 
from the secondary component to the more massive primary star. Assuming a conservative mass transfer, the transfer rate is 
5.72 $\times$ 10$^{-8}$ M$_\odot$ yr$^{-1}$. The observed value is smaller by a factor of about 60\% compared with 
the predicted rate of 1.43 $\times$ 10$^{-7}$ M$_\odot$ yr$^{-1}$ calculated by assuming that the secondary transfers 
its present mass to the primary on a thermal time scale. Thus, the parabolic variation might originate from 
non-conservative mass transfer. The result is consistent with a recent study by Yildiz \& Do\u{g}an (2013), finding 
that $\sim$ 34 \% of the mass from the secondary is transferred to the primary component and the remainder is lost 
from the binary system.

\section{DISCUSSION AND CONCLUSIONS}

In this paper, we have presented the physical nature and orbital behavior of DK Cyg derived from detailed studies of 
all available data. Historical light curves, including our own, indicate that the secondary minimum displays 
a total eclipse but is asymmetric and distorted. Further, the light maxima are shifted and the eclipses indicate 
clear evidence for short-time brightness disturbance. These features may be ascribed to surface inhomogeneities, which 
is satisfactorily modeled by a magnetic cool spot on the primary star. The modeled spot almost certainly corresponds to 
a spotted region rather than a single large spot. Our results show that the eclipsing system is a hot overcontact binary 
with a relatively small temperature difference of 489 K, unlike the previous values of Baran et al. (2004) and 
Elkhateeb \& Nouh (2014). From the computed absolute parameters, it is possible to consider the evolutionary state in 
M-R, M-L, and the Hertzsprung-Russell (HR) diagrams. The locations of the component stars in these diagrams do conform 
to the general pattern of W UMa binaries. The primary star lies between the ZAMS and the TAMS, while the secondary is 
oversized and overluminous for its mass in the first two diagrams and to the left of the main-sequence band on 
the HR diagram. This can be explained as a result of luminosity transfer from the primary to the secondary component 
(Kuiper 1948; Lucy 1968)

The 78-yr period modulation in the eclipse timing diagram can be caused by changes of an active star's internal 
angular momentum distribution as the star goes through a magnetic activity cycle (Applegate 1992, Lanza et al. 1998). 
But, the magnetic mechanism never displays a pattern of alternating period decreases and increases for systems with 
spectra earlier than about F5 (Hall 1989, Liao \& Qian 2010). This indicates that the Applegate model cannot explain 
the observed period modulations. On the other hand, eclipse times can be shifted from conjunction instants by 
asymmetrical eclipse minima originating from starspot activity and/or even by the method of measuring the timings of 
minimum (Tran et al. 2013; Lee et al. 2014b, 2015). The light-curves synthesis method developed by W-D can 
give better information for the conjunction instants than other methods. Because the three datasets of DK Cyg were modeled 
for spot parameters, we calculated a minimum epoch for each eclipse curve in these datasets with the W-D code by adjusting 
only the ephemeris epoch ($T_0$). The results are given in Table 8, together with the previously-calculated timings for 
comparison and the differences between the two values are much smaller than the observed amplitude (about 0.007 d) of 
the LTT variation. Therefore, the periodic oscillation most likely arises from the LTT effect due to 
a low-mass tertiary companion orbiting the inner eclipsing binary.

The existence of the third component in DK Cyg is consistent with the suggestion of Pribulla \& Rucinski (2006) that 
most W UMa-type binaries exist in multiple systems. The circumbinary companion may have played an important role in shrinking 
the primordial wide binary into the current configuration through Kozai oscillation (Kozai 1962; Pribulla \& Rucinski 2006) 
or a combination of the Kozai cycle and tidal friction (Fabrycky \& Tremaine 2007). The present overcontact pair will ultimately
coalesce into a rapid-rotating single star by angular momentum loss due to magnetic braking (Bradstreet \& Guinan 1994; 
Tylenda et al. 2011) and then the triple system will become a moderately wide binary star. Because only about 74 \% of 
the LTT period has been covered by the photoelectric and CCD data, precise long-term timing measurements are required 
to identify and understand the substellar companion proposed for the eclipsing system.

\acknowledgments{ }

The authors wish to thank the staffs of SOAO and LOAO for assistance during our observations. We appreciate the careful reading 
and valuable comments of the anonymous referee. This research has made use of the Simbad database maintained at CDS, Strasbourg, 
France. We have used data from the WASP public archive in this research. The WASP consortium comprises of the University of 
Cambridge, Keele University, University of Leicester, The Open University, The Queen's University Belfast, St. Andrews University 
and the Isaac Newton Group. Funding for WASP comes from the consortium universities and from the UK's Science and 
Technology Facilities Council. This work was supported by the KASI grant 2015-1-850-04. M.W. acknowledges support by 
the Research Program MSM0021620860 {\it Physical Study of objects and processes in the Solar System and in Astrophysics} of 
the Ministry of Education of the Czech Republic.

\clearpage
\begin{figure}
 \includegraphics[]{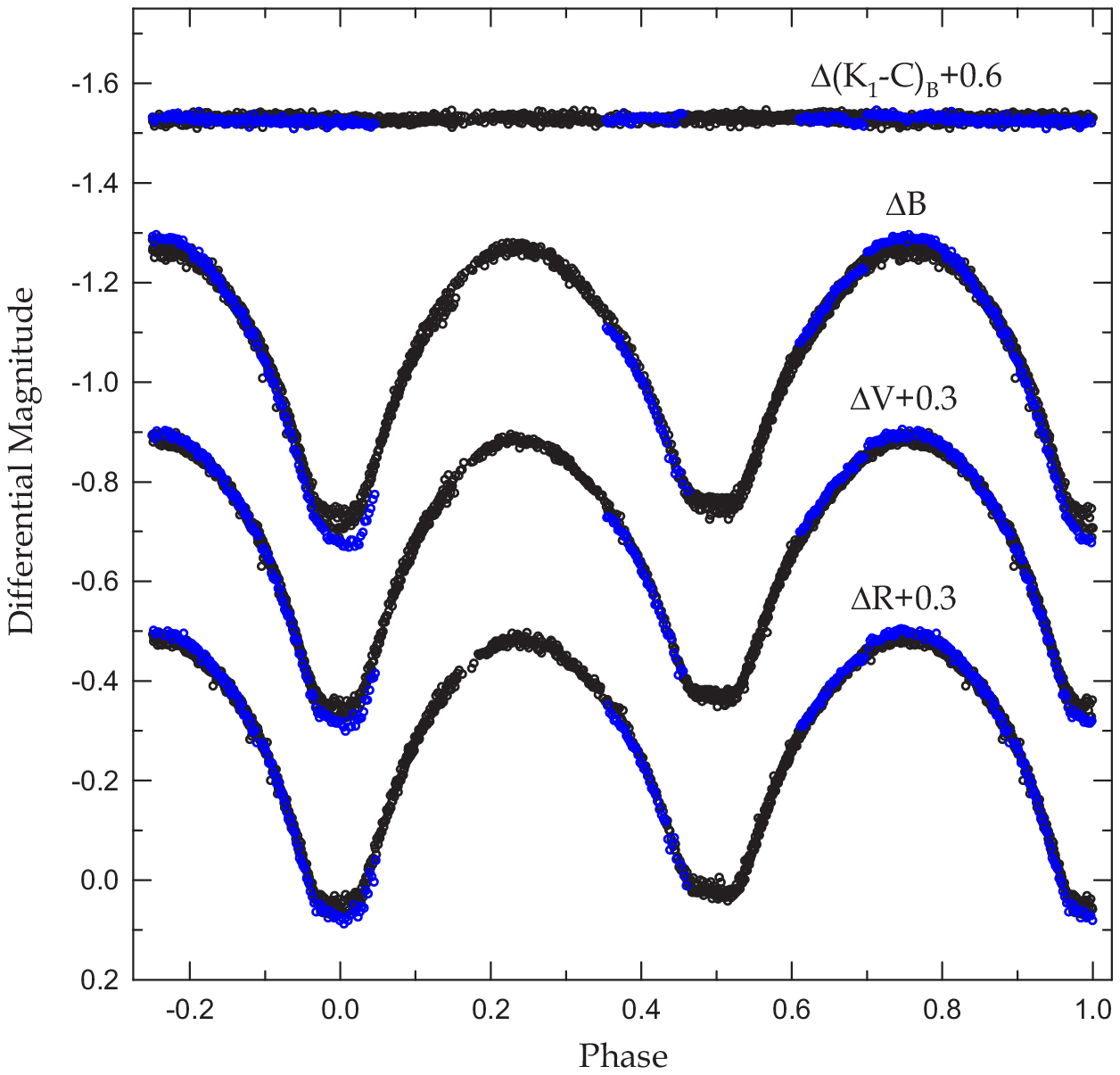}
 \caption{$BVR$ light curves of DK Cyg observed at SOAO. The uppermost $\Delta$(K$_1$--C)$_{\rm B}$ is the magnitude differences 
 between the check and comparison stars in the $B$ bandpass. Blue circles are the measurements on 2012 September 23 }
 \label{Fig1}
\end{figure}

\begin{figure}
 \includegraphics[]{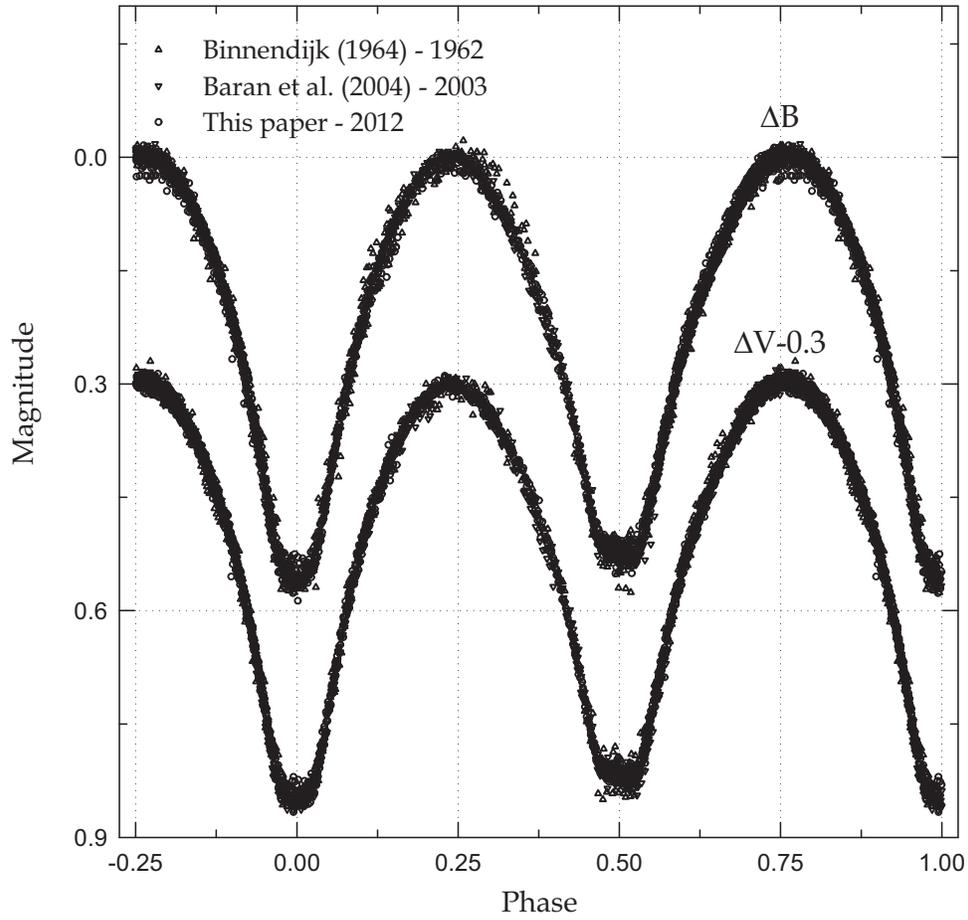}
 \caption{Composite $BV$ light curve of DK Cyg obtained from 1962 to 2012. The observations have been made by requiring 
 the maximum lights at the first quadrature to be identical (i.e. 0.0 mag). }
\label{Fig2}
\end{figure}

\begin{figure}
 \includegraphics[]{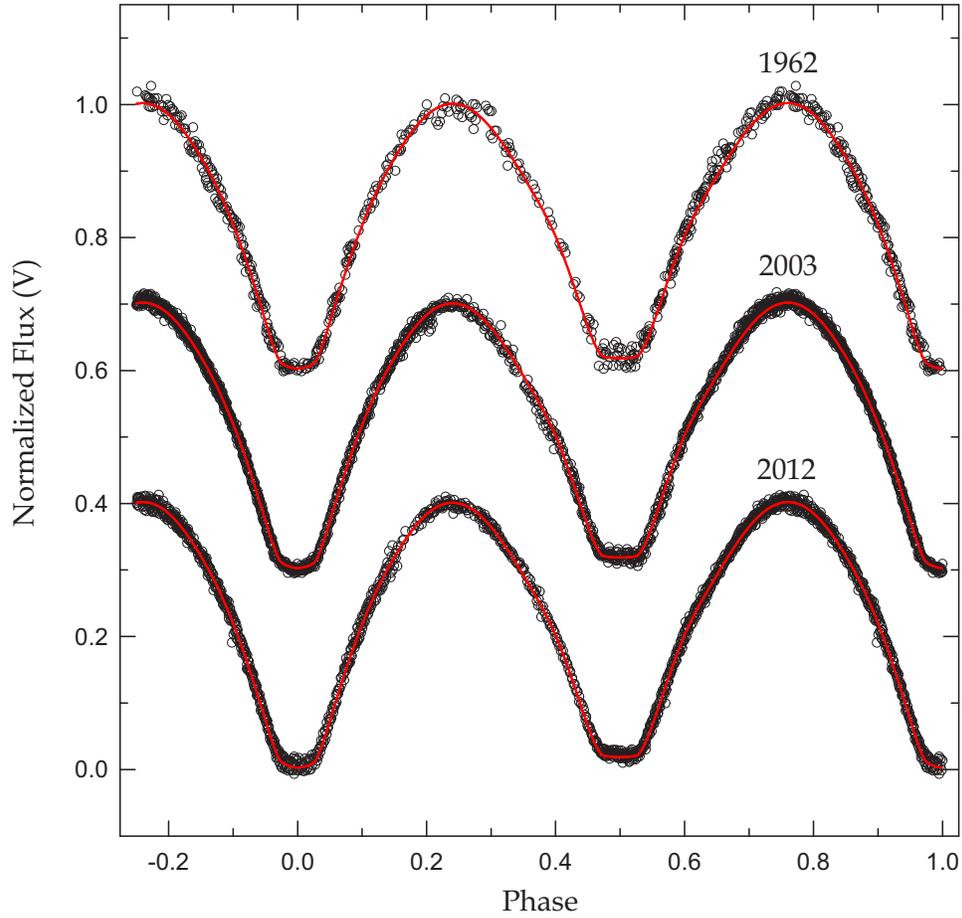}
 \caption{Normalized $V$ observations with fitted model light curves. The light curves of 2003 and 2012 are displaced vertically 
 for clarity. The continuous curves represent the solutions obtained from the cool-spot model on the primary star listed in Table 3. }
 \label{Fig3}
\end{figure}

\begin{figure}
 \includegraphics[]{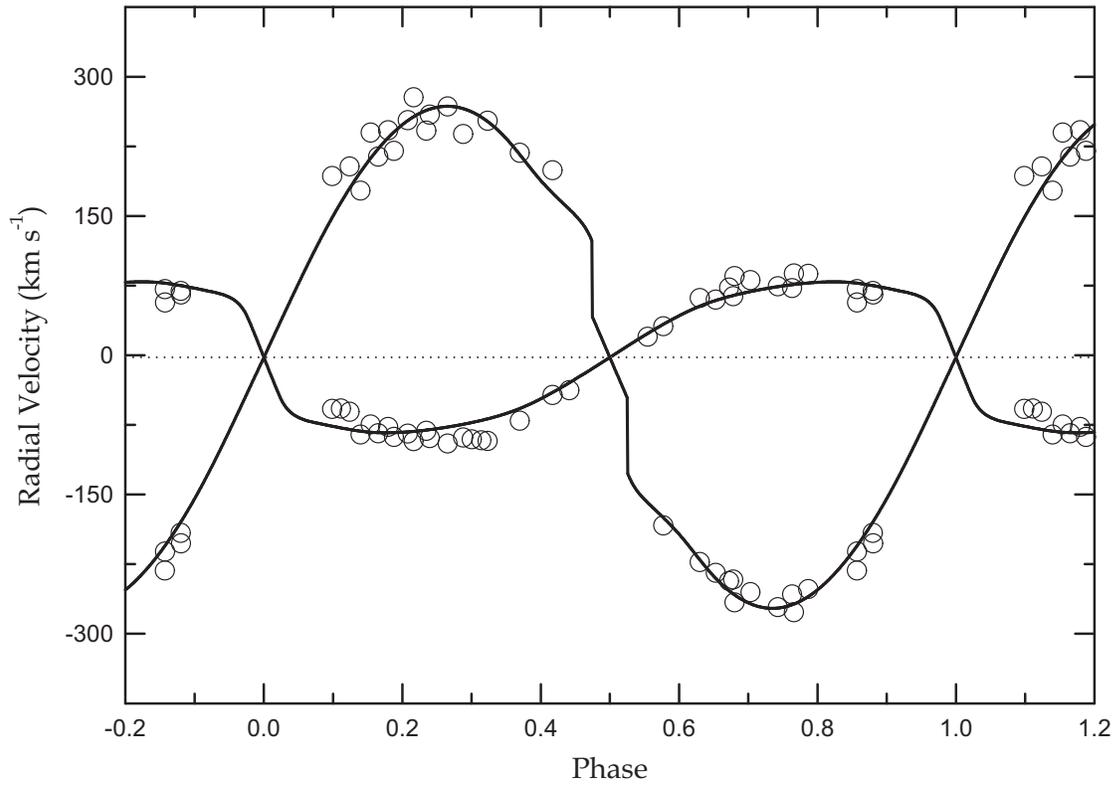}
 \caption{Radial-velocity curves of DK Cyg. The open circles are the measurements of Rucinski \& Lu (1999), while the solid curves denote
 the result from consistent light and velocity curve analysis. The dotted line refers to the systemic velocity of $-$2.2 km s$^{-1}$.}
\label{Fig4}
\end{figure}

\begin{figure}
 \includegraphics[]{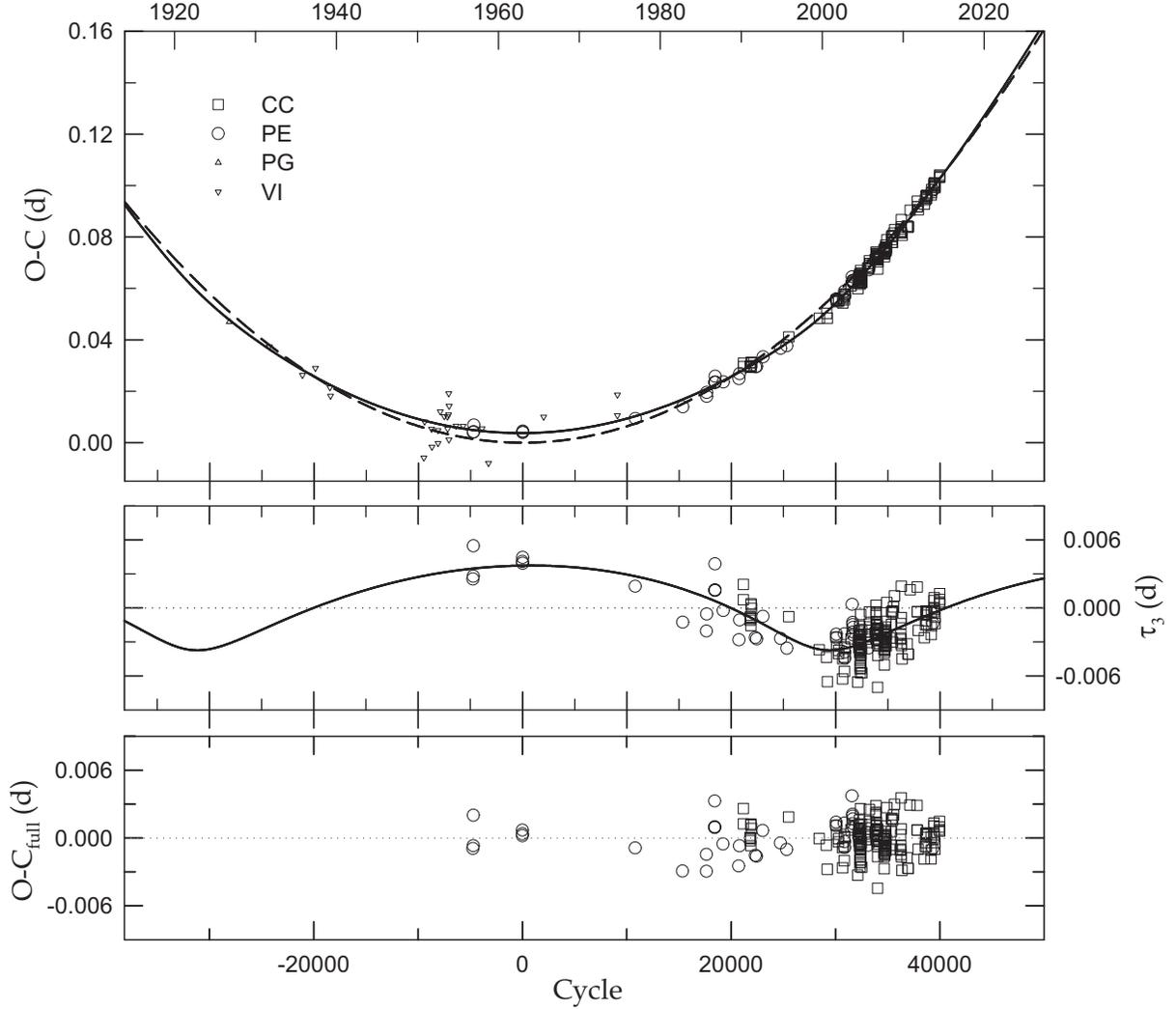}
 \caption{In the top panel the $O$--$C$ diagram of DK Cyg is constructed with the linear terms of the quadratic {\it plus} LTT ephemeris.
 The full ephemeris is drawn as the solid curve and the dashed parabola is only due to the quadratic term of Equation (1). 
 CC, PE, PG, and VI stand for CCD, photoelectric, photographic, and visual minima, respectively. The middle panel refers to 
 the LTT orbit ($\tau_3$) and the bottom panel shows the photoelectric and CCD residuals from the complete ephemeris. }
 \label{Fig5}
\end{figure}

\clearpage
\begin{deluxetable}{crcrcr}
\tabletypesize{\small}
\tablewidth{0pt} 
\tablecaption{CCD photometric observations of DK Cyg observed at SOAO.}
\tablehead{
\colhead{HJD} & \colhead{$\Delta B$} & \colhead{HJD} & \colhead{$\Delta V$} & \colhead{HJD} & \colhead{$\Delta R$} 
}
\startdata
2,456,190.93360 & $-$0.7223  &  2,456,190.93400 & $-$0.0438  &  2,456,190.93435 &     0.3549   \\
2,456,190.93497 & $-$0.7185  &  2,456,190.93748 & $-$0.0531  &  2,456,190.93785 &     0.3375   \\
2,456,190.93859 & $-$0.7368  &  2,456,190.93903 & $-$0.0527  &  2,456,190.93940 &     0.3488   \\
2,456,190.94015 & $-$0.7310  &  2,456,190.94055 & $-$0.0594  &  2,456,190.94090 &     0.3495   \\
2,456,190.94164 & $-$0.7297  &  2,456,190.94204 & $-$0.0627  &  2,456,190.94239 &     0.3327   \\
2,456,190.94313 & $-$0.7482  &  2,456,190.94354 & $-$0.0652  &  2,456,190.94388 &     0.3301   \\
2,456,190.94463 & $-$0.7586  &  2,456,190.94503 & $-$0.0830  &  2,456,190.94538 &     0.3053   \\
2,456,190.94612 & $-$0.7751  &  2,456,190.94653 & $-$0.0937  &  2,456,190.94688 &     0.2940   \\
2,456,190.94762 & $-$0.7922  &  2,456,190.94801 & $-$0.1086  &  2,456,190.94833 &     0.2731   \\
2,456,190.94907 & $-$0.8071  &  2,456,190.94946 & $-$0.1280  &  2,456,190.94978 &     0.2631   \\
\enddata
\tablecomments{This table is available in its entirety in machine-readable and Virtual Observatory (VO) forms 
in the online journal. A portion is shown here for guidance regarding its form and content.}
\end{deluxetable}

\begin{deluxetable}{lccc}
\tablewidth{0pt}
\tablecaption{Radial velocity and light-curve sets for DK Cyg.}
\tablehead{
\colhead{Reference}          & \colhead{Season} & \colhead{Data type} & \colhead{$\sigma$$\rm ^a$} }
\startdata
Rucinski \& Lu (1999)        & 1996$-$1997      & RV1                 & 11.0 km s$^{-1}$   \\
                             &                  & RV2                 & 13.3 km s$^{-1}$   \\
Binnendijk (1964)            & 1962             & $B$                 & 0.0155             \\
                             &                  & $V$                 & 0.0101             \\
Baran et al. (2004)          & 2003             & $B$                 & 0.0059             \\
                             &                  & $V$                 & 0.0064             \\
                             &                  & $R$                 & 0.0068             \\
SOAO                         & 2012             & $B$                 & 0.0085             \\
                             &                  & $V$                 & 0.0058             \\
                             &                  & $R$                 & 0.0062             \\
\enddata
\tablenotetext{a}{For the light curves, in units of total light at phase 0.25.}
\end{deluxetable}

\begin{deluxetable}{lcc}
\tablewidth{0pt}
\tablecaption{Binary parameters of DK Cyg$\rm ^a$.}
\tablehead{
\colhead{Parameter}                     & \colhead{Primary} & \colhead{Secondary}
}
\startdata                                                                         
$T_0$ (HJD)                             & \multicolumn{2}{c}{2,437,999.58029(8)}    \\
$P$ (day)                               & \multicolumn{2}{c}{0.470690658(7)}        \\
d$P$/d$t$ (10$^{-10}$)                  & \multicolumn{2}{c}{2.681(6)}              \\
$\gamma$ (km s$^{-1}$)                  & \multicolumn{2}{c}{$-$2.2(1.7)}           \\
$a$ (R$_\odot$)                         & \multicolumn{2}{c}{3.40(4)}               \\
$q$                                     & \multicolumn{2}{c}{0.307(8)}              \\
$i$ (deg)                               & \multicolumn{2}{c}{82.75(6)}              \\
$T$ (K)                                 & 7500(200)         & 7011(200)             \\
$\Omega$                                & 2.396(1)          & 2.396                 \\
$\Omega_{\rm in}$                       & \multicolumn{2}{c}{2.481}                 \\
$X$, $Y$                                & 0.658, 0.230      & 0.641, 0.258          \\
$x_{B}$, $y_{B}$                        & 0.605(5), 0.312   & 0.775(21), 0.294      \\
$x_{V}$, $y_{V}$                        & 0.597(5), 0.285   & 0.649(18), 0.295      \\
$x_{R}$, $y_{R}$                        & 0.588(6), 0.266   & 0.571(16), 0.296      \\
$L/(L_1+L_2)_{B}$                       & 0.8124(19)        &  0.1876               \\
$L/(L_1+L_2)_{V}$                       & 0.7917(12)        &  0.2083               \\
$L/(L_1+L_2)_{B}$                       & 0.8124(6)         &  0.1876               \\
$L/(L_1+L_2)_{V}$                       & 0.7917(6)         &  0.2083               \\
$L/(L_1+L_2)_{R}$                       & 0.7739(6)         &  0.2261               \\
$L/(L_1+L_2)_{B}$                       & 0.8124(7)         &  0.1876               \\
$L/(L_1+L_2)_{V}$                       & 0.7917(5)         &  0.2083               \\
$L/(L_1+L_2)_{R}$                       & 0.7739(5)         &  0.2261               \\
$r$ (pole)                              & 0.4720(3)         & 0.2821(5)             \\
$r$ (side)                              & 0.5124(5)         & 0.2968(7)             \\
$r$ (back)                              & 0.5444(7)         & 0.3484(15)            \\
$r$ (volume)$\rm ^b$                    & 0.5112            & 0.3100                \\ [1.0mm]
\multicolumn{3}{l}{Spot parameters:}                                                \\ 
Colatitude (deg)                        & 75.5(4)           & \dots                 \\
Longitude (deg)                         & 181.7(2)          & \dots                 \\
Radius (deg)                            & 33.72(6)          & \dots                 \\
$T$$\rm _{spot}$/$T$$\rm _{local}$      & 0.942(1)          & \dots                 \\
$\Sigma W(O-C)^2$                       & \multicolumn{2}{c}{0.0129}                \\[1.0mm]
\enddata
\tablenotetext{a}{Bandpass luminosities are listed in the same order as entries in Table 2. }
\tablenotetext{b}{Mean volume radius.}
\end{deluxetable}

\begin{deluxetable}{lccc}
\tablewidth{0pt}
\tablecaption{Spot and luminosity parameters for each dataset.}
\tablehead{
\colhead{Parameter}                       & \colhead{Binnendijk}  & \colhead{Baran et al.}  & \colhead{This paper}               
}                                                                                                                                                                      
\startdata                                                                                                                                                             
$T_0$ (HJD)$\rm ^a$                       & 37,999.58429(10)      & 52,888.530483(23)       & 56,195.165660(35)    \\
$P$ (day)                                 & 0.47070652(94)        & 0.47068419(70)          & 0.47069557(95)       \\
Colatitude$_1$ (deg)                      & 87.3(7.3)             & 81.65(31)               & 75.52(21)            \\
Longitude$_1$ (deg)                       & 178.95(77)            & 181.34(24)              & 181.97(30)           \\
Radius$_1$ (deg)                          & 33.57(24)             & 33.81(10)               & 33.72(10)            \\
$T$$\rm _{spot,1}$/$T$$\rm _{local,1}$    & 0.940(1)              & 0.942(1)                & 0.945(1)             \\
$L_1/(L_1+L_2)_{B}$                       & 0.8124(7)             & 0.8124(2)               & 0.8129(3)            \\
$L_1/(L_1+L_2)_{V}$                       & 0.7917(5)             & 0.7917(2)               & 0.7919(2)            \\
$L_1/(L_1+L_2)_{R}$                       & \dots                 & 0.7739(2)               & 0.7739(2)            \\
\enddata
\tablenotetext{a}{HJD 2,400,000 is suppressed.}
\end{deluxetable}

\begin{deluxetable}{lcc}
\tablewidth{0pt}
\tablecaption{Absolute parameters for DK Cyg.}
\tablehead{
\colhead{Parameter}    & \colhead{Primary}    & \colhead{Secondary}}
\startdata                                    
$M$ (M$_\odot$)                         & 1.82(7)          & 0.56(2)              \\
$R$ (R$_\odot$)                         & 1.74(3)          & 1.05(2)              \\
$\log$ $g$ (cgs)                        & 4.22(2)          & 4.14(2)              \\
$\rho$ (g cm$^3)$                       & 0.49(3)          & 0.68(4)              \\
$L$ (L$_\odot$)                         & 8.5(9)           & 2.4(3)               \\
$M_{\rm bol}$ (mag)                     & $+$2.40(12)      & $+$3.78(13)          \\
BC (mag)                                & $+$0.03          & $+$0.03              \\
$M_{\rm V}$ (mag)                       & $+$2.37(12)      & $+$3.75(13)          \\
Distance (pc)                           & \multicolumn{2}{c}{366(21)}              
\enddata                                                     
\end{deluxetable}

\begin{deluxetable}{lllrrcl}
\tabletypesize{\small}
\tablewidth{0pt} 
\tablecaption{Observed photoelectric and CCD times of minimum light for DK Cyg.}
\tablehead{
\colhead{HJD} & \colhead{HJED$\rm ^a$} & Error & \colhead{Epoch} & \colhead{$O$--$C_{\rm full}$} & \colhead{Min} & \colhead{References}  \\
\colhead{(2,400,000+)} & \colhead{(2,400,000+)} & & & & &  }
\startdata
35,762.391    & 35,762.39151  &               &  $-$4,753.0  &  $-$0.00092  &  I   &  Hinderer (1960)                       \\
35,778.3947   & 35,778.39521  &               &  $-$4,719.0  &  $-$0.00069  &  I   &  Szafraniec (1962)                     \\
35,787.3405   & 35,787.34101  &               &  $-$4,700.0  &  $+$0.00199  &  I   &  Szafraniec (1962)                     \\
37,995.5831   & 37,995.58361  &               &       -8.5   &  $+$0.00036  &  II  &  Binnendijk (1964)                     \\
37,999.5838   & 37,999.58431  &               &        0.0   &  $+$0.00018  &  I   &  Binnendijk (1964)                     \\
38,000.5257   & 38,000.52621  &               &        2.0   &  $+$0.00070  &  I   &  Binnendijk (1964)                     \\
43,081.6367   & 43,081.63725  &               &    10,797.0  &  $-$0.00088  &  I   &  Paparo et al. (1985)                  \\
45,225.4019   & 45,225.40252  &               &    15,351.5  &  $-$0.00293  &  II  &  Braune et al. (1983)                  \\
46,300.4635   & 46,300.46414  &               &    17,635.5  &  $-$0.00295  &  II  &  Paparo et al. (1985)                  \\
46,303.5245   & 46,303.52514  &               &    17,642.0  &  $-$0.00145  &  I   &  Paparo et al. (1985)                  \\
46,676.3155   & 46,676.31614  &               &    18,434.0  &  $+$0.00096  &  I   &  Awadalla (1994)                       \\
46,676.5508   & 46,676.55144  &               &    18,434.5  &  $+$0.00092  &  II  &  Awadalla (1994)                       \\
46,679.3773   & 46,679.37794  &               &    18,440.5  &  $+$0.00326  &  II  &  Awadalla (1994)                       \\
46,680.3164   & 46,680.31704  &               &    18,442.5  &  $+$0.00098  &  II  &  Awadalla (1994)                       \\
46,680.5517   & 46,680.55234  &               &    18,443.0  &  $+$0.00093  &  I   &  Awadalla (1994)                       \\
47,051.4561   & 47,051.45674  &               &    19,231.0  &  $-$0.00053  &  I   &  H\"ubscher \& Lichtenknecker (1988)   \\
47,758.4348   & 47,758.43545  &               &    20,733.0  &  $-$0.00247  &  I   &  H\"ubscher et al. (1990)              \\
47,790.4437   & 47,790.44435  &               &    20,801.0  &  $-$0.00069  &  I   &  H\"ubscher et al. (1990)              \\
47,963.662    & 47,963.66266  & $\pm$0.001    &    21,169.0  &  $+$0.00260  &  I   &  Wolf et al. (2000)                    \\
47,963.896    & 47,963.89666  & $\pm$0.001    &    21,169.5  &  $+$0.00126  &  II  &  Wolf et al. (2000)                    \\
\enddata
\tablenotetext{a}{HJD in the terrestrial time (TT) scale.}
\tablecomments{This table is available in its entirety in machine-readable and Virtual Observatory (VO) forms in the online journal. 
A portion is shown here for guidance regarding its form and content.}
\end{deluxetable}

\begin{deluxetable}{lcc}
\tablewidth{0pt}
\tablecaption{Parameters for the quadratic {\it plus} LTT ephemeris of DK Cyg. }
\tablehead{
\colhead{Parameter}     &  \colhead{Values}                      &  \colhead{Unit}
}
\startdata                 
$T_0$                   &  2,437,999.58039$\pm$0.00025           &  HJED                  \\
$P$                     &  0.470690696$\pm$0.000000018           &  d                     \\
$A$                     &  +(6.439$\pm$0.029)$\times 10^{-11}$   &  d                     \\
$a_{12}\sin i_{3}$      &  0.650$\pm$0.046                       &  AU                    \\
$\omega$                &  259$\pm$7                             &  deg                   \\
$e$                     &  0.509$\pm$0.049                       &                        \\
$n  $                   &  0.01262$\pm$0.00059                   &  deg d$^{-1}$          \\
$T$                     &  2,422,961$\pm$1640                    &  HJED                  \\
$P_{3}$                 &  78.1$\pm$3.6                          &  yr                    \\
$K$                     &  0.00374$\pm$0.00026                   &  d                     \\
$f(M_{3})$              &  0.0000451$\pm$0.0000038               &  $M_\odot$             \\
$M_3 \sin i_{3}$        &  0.065$\pm$0.003                       &  $M_\odot$             \\
$dP$/$dt$               &  +(9.993$\pm$0.046)$\times 10^{-8}$    &  d yr$^{-1}$           \\[1.0mm]
$\sigma _{\rm all} ^a$  &  0.0023                                &                        \\
$\sigma _{\rm pc} ^b$   &  0.0013                                &                        \\
$\chi^2 _{\rm red}$     &  1.054                                 &                        \\
\enddata
\tablenotetext{a}{rms scatter of all residuals.}
\tablenotetext{b}{rms scatter of the photoelectric and CCD residuals.}
\end{deluxetable}

\begin{deluxetable}{ccccccl}
\tablewidth{0pt}
\tablecaption{Minimum timings determined by the W-D code from individual eclipses.}
\tablehead{
\colhead{Observed$\rm^{a,b}$} & \colhead{W-D$\rm^{b}$} & \colhead{Error$\rm^{c}$} & \colhead{Difference$\rm^{d}$} & \colhead{Filter} & \colhead{Min} & References
}
\startdata
37,995.58361  &  37,995.58345  &  $\pm$0.00023  &  $+$0.00016  &  $BV$    &  II  &  Binnendijk     \\
37,999.58431  &  37,999.58417  &  $\pm$0.00013  &  $+$0.00014  &  $BV$    &  I   &  Binnendijk     \\
38,000.52621  &  38,000.52559  &  $\pm$0.00004  &  $+$0.00062  &  $BV$    &  I   &  Binnendijk     \\
52,863.35105  &  52,863.34966  &  $\pm$0.00005  &  $+$0.00139  &  $BVR$   &  II  &  Baran et al.   \\
52,888.53161  &  52,888.53135  &  $\pm$0.00004  &  $+$0.00044  &  $BVR$   &  I   &  Baran et al.   \\
52,898.41538  &  52,898.41494  &  $\pm$0.00004  &  $+$0.00044  &  $BVR$   &  I   &  Baran et al.   \\
52,903.35808  &  52,903.35811  &  $\pm$0.00004  &  $-$0.00003  &  $BVR$   &  II  &  Baran et al.   \\
56,191.16528  &  56,191.16487  &  $\pm$0.00005  &  $+$0.00041  &  $BVR$   &  II  &  This article   \\
56,192.10700  &  56,192.10639  &  $\pm$0.00010  &  $+$0.00061  &  $BVR$   &  II  &  This article   \\
56,195.16658  &  56,195.16682  &  $\pm$0.00006  &  $-$0.00024  &  $BVR$   &  I   &  This article   \\
56,217.99529  &  56,217.99515  &  $\pm$0.00006  &  $+$0.00014  &  $BVR$   &  II  &  This article   \\
56,218.93658  &  56,218.93632  &  $\pm$0.00006  &  $+$0.00026  &  $BVR$   &  II  &  This article   \\
56,221.99616  &  56,221.99599  &  $\pm$0.00008  &  $+$0.00017  &  $BVR$   &  I   &  This article   \\
\enddata
\tablenotetext{a}{cf. Table 6.}
\tablenotetext{b}{HJED 2,400,000 is suppressed.}
\tablenotetext{c}{Uncertainties yielded by the W-D code.}
\tablenotetext{d}{Differences between columns (1) and (2).}
\end{deluxetable}

\end{document}